\documentclass[11pt]{article}
\usepackage[sfdefault]{carlito} 
\usepackage{geometry}
\usepackage{setspace}
\usepackage{titlesec}
\usepackage{hyperref}
\usepackage{graphicx}
\usepackage{amsmath}
\usepackage{bm}
\usepackage{wrapfig}
\usepackage{xcolor}
\usepackage{authblk}
\usepackage{braket}
\usepackage[font=small,labelfont=bf]{caption}

\usepackage{amssymb,bbm,braket,xspace}
\usepackage[utf8]{inputenc}
\newcommand{\mat}[1]{\mathbf{\underline{#1}}}
\newcommand{\CODE}[1]{\texttt{#1}\xspace}
\newcommand{\FHIaims}{\CODE{FHI-aims}}
\newcommand{\Vibes}{\CODE{FHI-vibes}}

\newcommand{\controlin}{\CODE{control.in}}
\newcommand{\aimd}{\textit{ai}MD\xspace}

\renewcommand{\vec}[1]{\mathbf{#1}}
\renewcommand{\t}[1]{\text{#1}}
\newcommand{\matrixf}[1]{\mat{#1}}
\newcommand{\ChapVibes}[0]{6.1}

\newcommand{\ChapMLIP}[0]{8.4}

\newcommand{\ChapFundamentals}[0]{1.1}

\newcommand{\ChapELPA}[0]{2.5}
\newcommand{\ChapElecTrans}[0]{7.2}
\newcommand{\ChapHeat}[0]{7.1}
\newcommand{\ChapPIMD}[0]{6.3}
\newcommand{\ChapMLDENS}[0]{8.5}
\hypersetup{
colorlinks=true
}

\titleformat{\section}{\normalfont\fontsize{11}{13}\bfseries\sffamily}{\thesection}{1em}{}
\titleformat{\subsection}{\normalfont\fontsize{11}{13}\bfseries\sffamily}{\thesubsection}{1em}{}
\titleformat{\title}{\normalfont\fontsize{14}{16}\bfseries\sffamily}{}{0em}{}

\title{Temperature-dependent Electronic Spectral Functions from Band-Structure Unfolding}
\author[1,2]{Jingkai Quan}
\author[1,3]{Min-Ye Zhang} 
\author[1,4]{Nikita Rybin} 
\author[1,5]{Marios Zacharias} 
\author[3]{Xinguo Ren} 
\author[6]{Hong Jiang}
\author[1]{Matthias Scheffler}
\author[1,7]{Christian Carbogno}
\affil[1]{The NOMAD Laboratory at the Fritz-Haber-Institut der Max-Planck-Gesellschaft \\ Faradayweg 4-6, 14195, Berlin, Germany}
\affil[2]{Max-Planck Institute for the Structure and Dynamics of Matter, Luruper Chausse 149, 22761, Hamburg, Germany}
\affil[3]{Institute of Physics, Chinese Academy of Sciences, 3rd South Str. 8, Beijing 100190, China}
\affil[4]{{\it Current address:} Skolkovo Institute of Science and Technology, Bolshoi bulvar 30, build.1, 121205, Moscow, Russia}
\affil[5]{{\it Current address:} Univ Rennes, INSA Rennes, CNRS, Institut FOTON - UMR 6082, F-35000 Rennes, France}
\affil[6]{Beijing National Laboratory for Molecular Sciences,  College of Chemistry and Molecular Engineering, Peking University, 100871 Beijing, China}
\affil[7]{{\it Current address:} Theory Department, Fritz Haber Institute of the Max Planck Society, Faradayweg 4-6, 14195 Berlin, Germany}
\date{}

\begin{document}

\maketitle

\section*{Summary} 
\begin{wrapfigure}{r}{0.5\textwidth}
  \centering
    \includegraphics[width=0.50\textwidth]{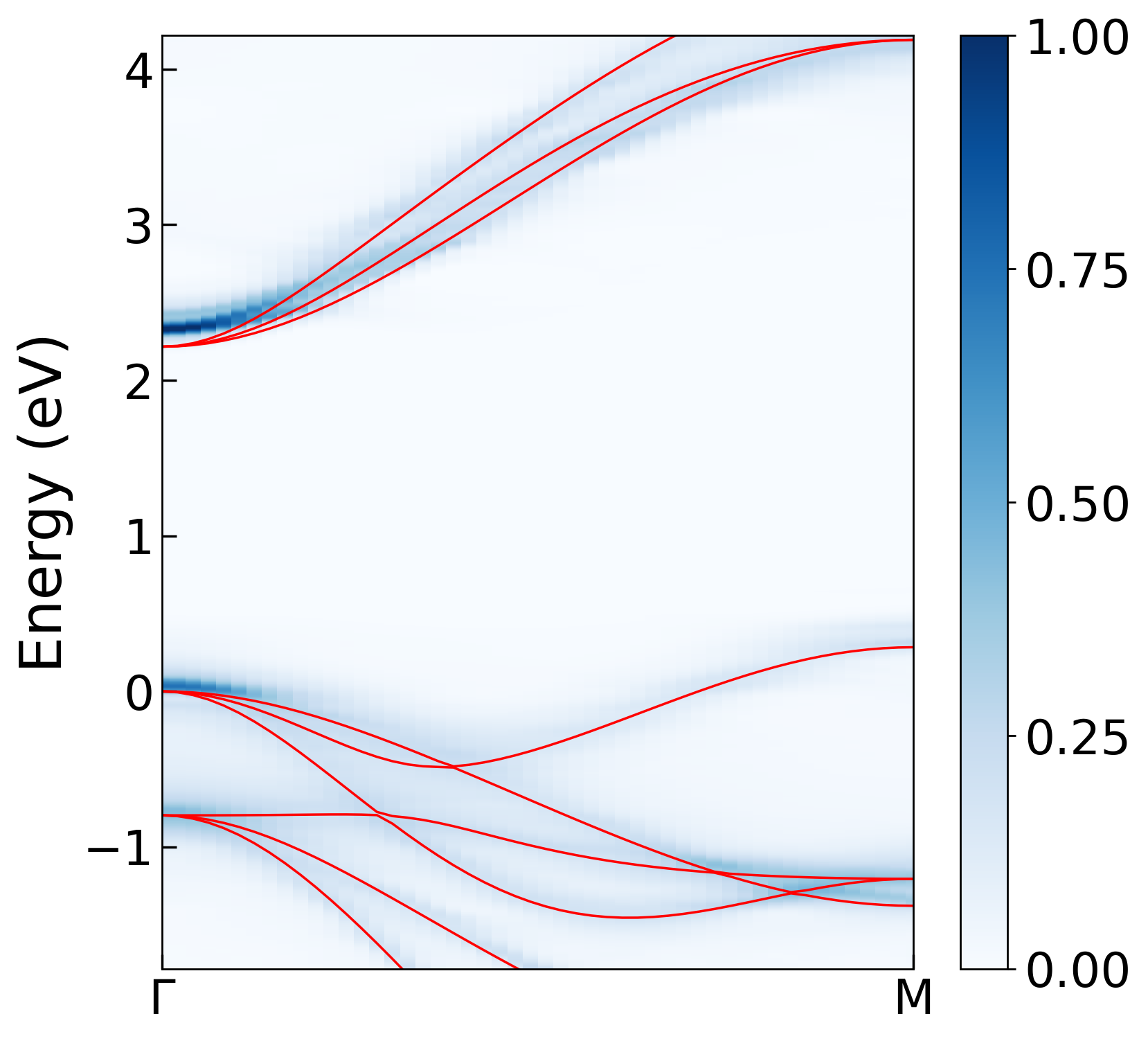}
    \caption{{Electronic spectral function at 300K and the static band structure (red lines) of SrTiO$_3$.}}
    \label{fig:Sketch1}
\end{wrapfigure}
The electronic band structure~$\varepsilon_n(\vec{k})$, which describes the periodic dependence of the electronic quantum states~$n$ on the lattice momentum~$\vec{k}$ 
in reciprocal space, 
is one of the fundamental concepts in solid-state physics. It is key to our understanding of solid-state devices, since it allows rationalizing the electronic 
properties of periodic materials,~e.g.,~to discern semiconductors with a direct band gap from those with an indirect one. In spite of that, the electronic band 
structure~$\varepsilon_n(\vec{k})$ is actually only well-defined for static nuclei,~i.e.,~immobile nuclei fixed at their crystallographic positions. This constitutes a severe approximation that does not even hold 
in the limit of zero Kelvin due to the quantum-nuclear zero-point motion. To account for these thermodynamics effects, the band-structure concept can be generalized
by introducing a temperature-dependent~($T$) spectral-function~$\braket{A({\bf k}, E)}_T$, as shown in Fig.~\ref{fig:Sketch1}. In this case, the electronic quantum
states at each reciprocal-vector $\vec{k}$ are no longer sharp values~$\varepsilon_n(\vec{k})$, but are characterized by a finite-width distribution. Several
fundamental physical properties and mechanisms can only be understood and computed when the coupling between nuclear and electronic degrees of freedom is 
accounted for. For instance, this includes the temperature-dependence of key electronic properties such as the band gap~\cite{Giustino.2010,Cannuccia.2012,Ponce.2014,Zacharias.2020},
optical absorption spectra~\cite{Noffsinger.2012,Zacharias.2015}, and electronic transport coefficients~\cite{Mustafa.2016,Zhou.2018,Ponce.2020}, also see Contrib.~\ChapElecTrans.

One possible route to compute spectral functions and the associated observables is many-body perturbation theory, a formally elegant and 
computationally efficient framework for treating electron-phonon coupling that is applicable within and beyond the Born-Oppenheimer approximation~\cite{Giustino.2017}. However, both the 
dynamics of the nuclei and that of the electrons are treated approximately in these methods by using the
harmonic approximation for the nuclear degrees of freedom, also see Contribs.~\ChapVibes and~\ChapHeat, and expressing the electronic response~(including
some second-order effects~\cite{Giustino.2017}) in terms of linear-order electron-phonon coupling elements. These approximations may well fail at elevated temperatures and/or for mobile atoms. For instance, it has been
demonstrated that the complete concept of phonons can break down in systems exhibiting spontaneous defect formations, even if these defects are 
short-lived~\cite{Knoop.2023yj}. To avoid potential inaccuracies from the aforementioned approximations, the electronic spectral function can be equally obtained 
in a non-perturbative fashion within the Born-Oppenheimer approximation, hence capturing higher-order couplings between electronic and vibrational degrees of freedom. 
To this end, {\it ab initio} molecular dynamics~(\aimd) can be run to accurately sample 
the full anharmonic potential-energy surface. The spectral-function is then obtained as the 
thermodynamic average~$\braket{A({\bf k}, E)}_T$ of the instantaneous electronic-band structures~$\varepsilon_n(\vec{k})$ observed during this 
dynamics~\cite{Zacharias.20204h9, Nery_Mauri_prb2022, Nikita_phd_thesis, BZU_arxiv}.
 
\begin{wrapfigure}{r}{0.45\textwidth}
  \centering
    \includegraphics[width=0.450\textwidth]{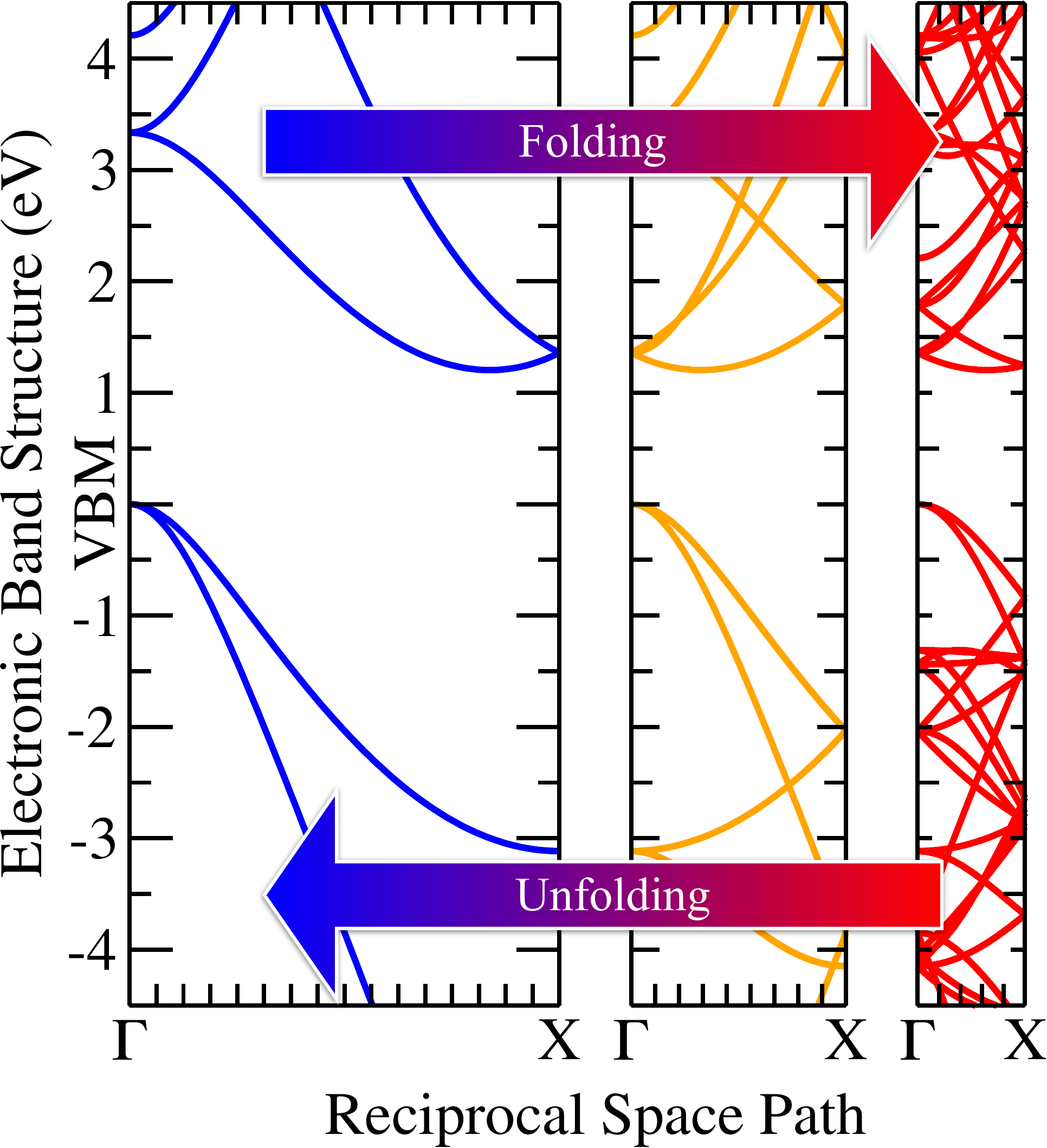}
    \caption{Folding viz. unfolding visualized for the HSE06 band structure of silicon along the high-symmetry $\Gamma\rightarrow X$~path for the primitive unit cell~(left), 
an eight-atom cubic conventional cell~(middle), and a 64-atom cubic supercell~(right).}
    \label{fig:Sketch2}
\end{wrapfigure}
From an electronic-structure point of view, one major hurdle in the implementation of the non-perturbative method arises from the fact that accurately sampling
the dynamics in solids requires extended supercells, see Contrib.~\ChapVibes. However, the electronic band-structures' topology is fundamentally connected 
to the translational symmetries of a crystal, which results in band-structure folding in supercell electronic-structure theory calculations. In this case, 
electronic states associated to different $\vec{k}$-vectors in the first Brillouin zone are mapped onto the same $\vec{K}$-vector in the reduced Brillouin zone 
when extended supercells are used in real space, see Fig.~\ref{fig:Sketch2}. Accordingly, the uttermost topological information is lost if the spectral function is computed as simple average
over folded supercell band structures. Rather, it is necessary to first recover the representation in the first Brillouin zone, a process often referred to as 
{\it unfolding}~\cite{Boykin.2005,Ku.2010,Popescu.2012,Allen.2013py8,suhuai_review} and visualized in Fig.~\ref{fig:Sketch2}.
In this contribution, we shortly describe the implementation of the band-structure unfolding technique in the electronic-structure theory package \FHIaims and the
updates made since its original development~\cite{Zacharias.20204h9}.

\section*{Current Status of the Implementation}
The main difficulty in the implementation of band-unfolding techniques in \FHIaims stems from the usage of a non-orthogonal, atomic-centered basis. In this numeric atomic-orbitals~(NAO) basis, 
atom displacements also affect the basis functions, so that changes in the basis set and in the overlap matrix need to be incorporated in the unfolding weight derivation~\cite{Zacharias.20204h9}. 
We discuss the unfolding technique by establishing a relationship between the electronic structure obtained for a primitive cell~(PC) 
and the one obtained for a supercell~(SC). Notationwise, properties associated to the PC or SC are denoted by using lower- and upper-case letters, 
respectively, and/or PC and SC subscripts when needed for clarity. Accordingly, the PC is characterized by the set of lattice vectors~$\matrixf{a} = \left( \vec{a}_1, \vec{a}_2, \vec{a}_3 \right)$ 
and the supercell~(SC) by the lattice vectors~$\matrixf{A} = \matrixf{a}\cdot\matrixf{M}$. Here, the lattice vectors $\bf a$ and $\bf A$ are column vectors, as by \FHIaims convention, and 
the $\matrixf{M}$ is a non-singular matrix with integer entries, 
implying that the volume of the SC is  $m=\left|\text{det}(\mathbf{\underline{M}})\right|$ times larger than that of the PC. Similarly, the lattice vectors~$\matrixf{b}$ of 
the first Brillouin zone~(BZ) associated with the PC and the ones of the reduced BZ associated with the SC are related via~$\matrixf{B}=\matrixf{M}^{-1}\matrixf{b}$.
In turn, the volume of the reduced BZ is $m$ times smaller than that of the first BZ, so that $m$ different $\vec{k}$-vectors in the first BZ zone are mapped onto 
one and the same $\vec{K}$-vector in the reduced BZ. The unfolding technique reverses this mapping and re-establishes a representation in the first BZ.

To perform the mapping, the projection operator
\begin{equation}
    \matrixf{P}_{\bf k} = \ket{{\bf k}}\bra{{\bf k}} 
\label{EQ_proj}
\end{equation}
is used. Here, $\ket{{\bf k}}$ are the eigenvectors with eigenvalue~$\exp(i \vec{k} \cdot \vec{a})$ that solve the eigenvalue 
problem\footnote{For the sake of clarity, the derivation here assumes non-degenerate eigenvalues. A discussion of the more general case including degeneracy can be found in Ref.~\cite{BZU_arxiv}.}
\begin{equation}
\mathbf{t}\ket{\bf k} = \exp(i \vec{k} \cdot \vec{a}) \ket{{\bf k}}
\label{EQ_eval}
\end{equation}
 for the translational operators~$\vec{t}$ associated to the lattice vectors~$\vec{a}$.
With that, it is possible to obtain the weights
\begin{equation} 
 W^{\bf k}_{{\bf K}N} = \braket{ \Psi_{{\bf K}N} | \matrixf{P}_{\bf k} | \Psi_{{\bf K}N}} = |\braket{{\bf k}|\Psi_{{\bf K}N}}|^2 \,,
\end{equation}
i.e.,~the contribution stemming from the subspace spanned by~$\ket{{\bf k}}$ associated with the translational symmetry of the PC to the SC wave function~$\Psi_{{\bf K}N}$.
For a ``perfect'' SC,~i.e.,~one consisting of identical PC replicas, these weights are either 0 or 1, and one can hence reconstruct the exact band structure in the first BZ 
from a SC calculation. For a disturbed SC, the weights~$W^{\bf k}_{{\bf K}N}$ become fractional, since the SC system is not invariant under PC-translations~$\vec{t}$.
Accordingly, the desired remapping can be obtained by summing over all SC wave functions~$\Psi_{{\bf K}N}$ and computing the electronic spectral function
\begin{eqnarray} \label{eq:spectral function}
    A({\bf k}, E) =   \sum_{ {\bf K}N} W^{\bf k}_{{\bf K}N}  \delta (E - E_{{\bf K}N}) \, .
\end{eqnarray}

In practice, the unfolding implementation in \FHIaims start from the representation of the wave function~$\Psi_{{\bf K}N} = \sum_i C_{Ni}(\vec{K}) \chi_i(\vec{K})$ in the SC,~where
 $C_{Ni}(\vec{K})$ denotes a Kohn-Sham expansion coefficient and $\chi_i(\vec{K})$ is a Bloch-like basis function build from numeric atomic orbitals. More details on these definition and notation can be found in 
Contrib.~\ChapFundamentals. In this SC Bloch-like basis, the algebraic representation of the PC translational operator is then constructed  via $t_{ij} = \braket{\chi_i(\vec{K})|\vec{t}|\chi_j(\vec{K})}$. 
Let us emphasize that these functions are not orthogonal in real space,~i.e.,~the overlap matrix~$S_{ij}(\vec{K}) = \braket{\chi_i(\vec{K})|\chi_j(\vec{K})}\neq \delta_{ij}$ is not diagonal. 
Accordingly, Eq.~(\ref{EQ_eval}) becomes a generalized eigenvector  problem in this representation, also see Contrib.~\ChapELPA. Its solution yields the 
eigenvalues~$\exp(i \vec{k} \cdot \vec{a})$,~i.e.,~those set of $\vec{k}$-points in the first BZ that this $\vec{K}$-point in the reduced BZ can be mapped to, and 
the set of eigenvectors~$\vec{k} = \sum_i F_i({\bf k})\chi_i(\vec{K})$ that are needed for constructing the projection operator $\matrixf{P}_{\bf k}$ defined in Eq.~(\ref{EQ_proj}). 
With that one obtains the following formula for the weight
\begin{eqnarray} \label{eq:unfold nonorthogonal}
    W^{\bf k}_{{\bf K}N} = |\mathbf{F}^{\dagger}(\vec{k}) \matrixf{S}_{\bf K}\mathbf{C}_{N}(\vec{K}) |^2  = |\mathbf{F}'^{\ \dagger}(\vec{k}) \mathbf{C}'_{N}(\vec{K}) |^2 \ .
\end{eqnarray}
In the last step, we have introduced the orthogonal representation $\mathbf{C}'_{N}(\vec{K}) = \matrixf{S}^{1 / 2}_{\bf K} \mathbf{C}_{N}(\vec{K})$ and $\mathbf{F}'(\vec{k}) = \matrixf{S}^{1 / 2}_{\bf K} \mathbf{F}(\vec{k})$ via a L{\"o}wedin transformation. Although the square root calculation~$\matrixf{S}^{1 / 2}_{\bf K}$ constitutes a computational overhead, the latter orthogonal representation is
advantageous, since this allows to find analytical expressions of the eigenvector expansion coefficients $\mathbf{F}'({\bf k})$ at each $\vec{K}$-point in the reduced BZ~\cite{BZU_arxiv}. With that
the numerical solution of the eigenvalue problem in Eq.~(\ref{EQ_eval}) is circumvented, leading to an overall speed-up of the procedure. Eventually, let us note that 
translations in three-dimensional systems form an Abelian group, so that the translations~$\vec{t}_\alpha$ associated to different lattice vectors~$\vec{a}_\alpha$ can be tackled
consecutively.

\section*{Usability and Tutorials} 
\begin{wrapfigure}{r}{0.45\textwidth}
  \centering
    \includegraphics[width=0.45\textwidth]{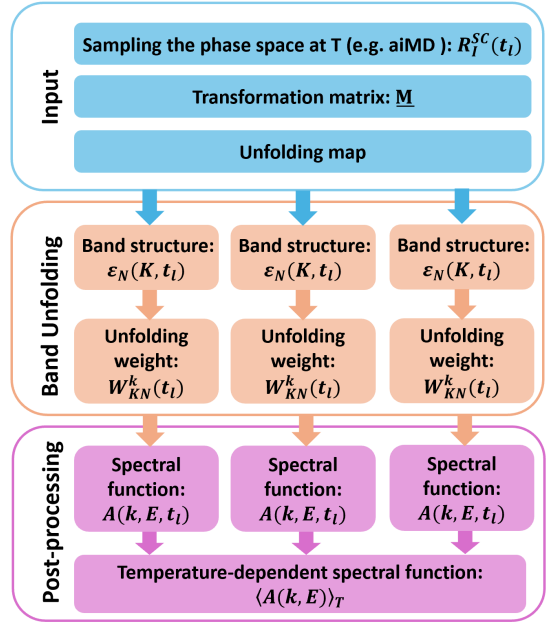}
    \caption{Schematics of a typical workflow used to obtain temperature-dependent spectral functions~$\braket{A({\bf k}, E)}_T$.}
    \label{FIG_FLOW}
\end{wrapfigure}
The overall workflow for computing temperature-dependent spectral functions~$\braket{A({\bf k}, E)}_T$ is sketched in Fig.~\ref{FIG_FLOW}. First,
one or multiple \aimd runs are performed in a SC at temperature~$T$ to sample the thermodynamic phase-space. From these trajectories, $L$~single, 
uncorrelated ``samples'',~i.e.,~atomic configurations~$\vec{R}_I^\t{SC}(t_l)$ at sufficiently distant times~$t_l$, are extracted and their 
electronic band-structure~$\varepsilon_N(\vec{K},t_l)$ is computed at a sufficiently dense~$\vec{K}$-grid. In this step, the unfolding routines
are used to map back the band-structure~$\varepsilon_N(\vec{K},t_l)$ onto the first BZ,~i.e.,~the weights~$W^{\bf k}_{{\bf K}N}(t_l)$ are computed
and stored on file together with $\varepsilon_N(\vec{K},t_l)$. In the last steps, the outputs are post-processed by computing the spectral 
functions~$A({\bf k}, E,t_l)$ of the individual samples, see Eq.~(\ref{eq:spectral function}), and by eventually computing the thermodynamic average
\begin{equation}
\braket{A({\bf k}, E)}_T = \frac{1}{L}\sum_{l=1}^L A({\bf k}, E,t_l) \;.
\end{equation}
A more detailed description including scripts to pre- and post-process the calculations and to analyze the data are given in the tutorial at~{ \href{https://fhi-aims-club.gitlab.io/tutorials/band-unfolding}{https://fhi-aims-club.gitlab.io/tutorials/band-unfolding}}. In this context, let us note that the method is not strictly restricted to \aimd, but any kind of sampling method can be used 
to explore the relevant phase-space. For instance, path-integral MD, see Contrib.~\ChapPIMD can be used instead, if it is necessary and desirable
to account for quantum-nuclear effects. Along this lines, also more approximative methods,~e.g.,~harmonic sampling as implemented in \Vibes~\cite{Knoop.2020fmh} 
and described in Contrib.~\ChapVibes or MD on machine-learned interatomic potentials as described in Contrib.~\ChapMLIP, can in principle be used. 
In such cases, however, correct spectral functions  can only be obtained if the employed approximations hold, as discussed in the respective contributions.

In more detail, the unfolding procedure in \FHIaims is part of the native band-structure post-processing tool. Accordingly it is invoked by 
 using the keyword {\tt output band} to define the reciprocal-space path for the SC,~i.e.,~the $\vec{K}$-points that shall be targeted and
by setting the keyword {\tt bs\_unfolding} in the \controlin file. Additionally, two more input files are required: 
{\tt transformation\_matrix.dat}, where one defines the transformation matrix $\matrixf{M}$ between PC lattice vector $\matrixf{a}$ and SC lattice vector $\matrixf{A}$,
and {\tt unfolding\_map.dat}, describing the mapping between atoms in the PC and the SC that is later used for constructing the translational operator~$\vec{t}$. The 
latter is an index map~$I\rightarrow j$ that relates the atomic coordinates in an unperturbed supercell~$\vec{R}_I^\t{SC} = \vec{R}_j^\t{PC} + \sum_\alpha n_\alpha \vec{a}_\alpha \quad n_\alpha \in \mathbb{Z}$ to 
the ones of the atoms~$\vec{R}_j^\t{PC}$ in a primitive cell. Upon completion, the unfolding weights for each SC $\bf K$-point are written into separate files named {\tt unfold\_k\_\#\#\#.out}
using the same \CODE{NXY} format used in the standard band structure output.

Currently, the band-unfolding implementation in \FHIaims supports all Bravais lattice, all exchange-correlation functionals, and all type of transformation matrices $\matrixf{M}$,
including non-diagonal ones typically needed to map primitive fcc and bcc structures to conventional cubic supercells. Both support for {\tt LAPACK} and {\tt ScaLAPACK} is implemented,
allowing for a trivially parallel parallelization over $\vec{K}$-points in the case of small systems requiring many $\vec{K}$-points~({\tt LAPACK}) and for a block cyclic distribution 
and distributed linear algebra in the case of large systems with few $\vec{K}$-points~({\tt ScaLAPACK}). With that, the band unfolding implementation in \FHIaims can routinely handle large 
supercells with small computational overhead. For instance, the unfolding only requires approx.~20\% of the total runtime when computing a 4,096-atom Silicon with one k-point, $> 100,000$ basis 
functions and $> 40,000$ states at the semi-local level of exchange-correlation. 

\begin{wrapfigure}{r}{0.5\textwidth}
	\centering
	\includegraphics[width=0.48\textwidth]{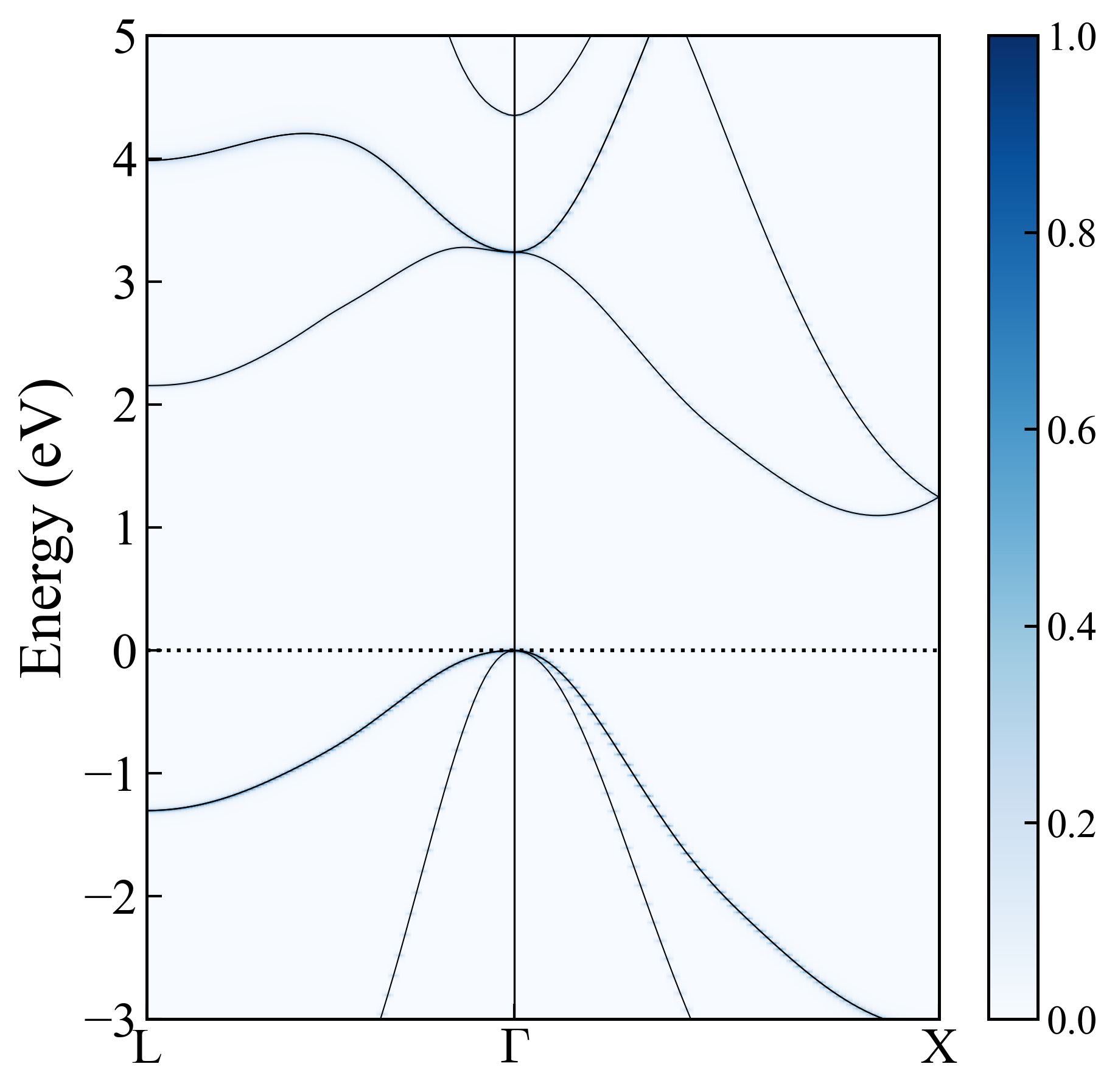}
	\caption{\label{fig:GW} $G_0W_0$ spectral function~(blue) for a perfect $2\times 2\times 2$ silicon super cell unfolded onto the first Brillouin zone. The $G_0W_0$ band structure obtained in a PC is shown as comparison in black.}
\end{wrapfigure}
Eventually, let us note that also {\it electron-electron} coupling,~i.e.,~electronic many-body effects, can play an important role for 
spectral functions~\cite{Antonius.2014,Karsai.2018}. To this end, the current implementation also supports the unfolding of~$G_0W_0$ calculations. In this case, the complex-valued self-energy is treated as a correction to the orbital energy, while the wave function is assumed to remain as in the Kohn-Sham scheme. The unfolded spectral function can then be computed in close analogy to Eq.~(\ref{eq:spectral function}). The required
weights are obtained at the Kohn-Sham level, while the delta-function~$\delta(E-E_{\vec{K}N})$ is replaced by the spectral function obtained at the $G_0W_0$ level. By this means, electron correlation is explicitly accounted for. As an example, Fig.~\ref{fig:GW} 
shows the unfolded $G_0W_0$ band structure for a perfect $2\times 2\times 2$ silicon super cell.

\section*{Future Plans and Challenges}

As described in this contribution, the band-structure unfolding procedures implemented in \FHIaims enable a routine assessment of temperature-dependent spectral functions both at the
density-functional theory and at the $G_0W_0$ level. From
a workflow perspective, the presented methodology is independent on the actual methodologies used (i)~for accurately sampling the thermodynamic phase-space and (ii)~for obtaining
the electronic band-structures~$\varepsilon_N(\vec{K})$ in the SC, as detailed above. With that, the availability of (sufficiently accurate) machine-learned interatomic potentials~\cite{Behler.2021} and 
of (sufficiently accurate) machine-learned electronic-structure theory~\cite{Li.2022,Zhang.2022,Lewis.2021}, see Contribs.~\ChapMLIP and \ChapMLDENS, can give access to much larger system sizes. 
Performance-wise, already the current implementation can handle such cases. However, access to such larger system sizes also enables the study of new physical questions,~e.g.,~the influence of point defects, 
of grain boundaries, and of solid-solid interfaces. While the overall theory holds also in such cases, the construction of the translational $\vec{t}$ operator will requires case-specific
adaptions to reflect the different types of breaking in translational symmetry.

Furthermore, the above-mentioned machine-learning methods might even help to solve a more fundamental open question with respect to non-adiabatic effects~\cite{Lazzeri.2006}, 
which can play a fundamental role for spectral functions, as evidence from many-body perturbation theory shows~\cite{Miglio.2020}. So far, however, such kind of effects could
not be studied with the herein presented non-perturbative approach, since it would require prohibitively expensive time-dependent electronic-structure theory
calculations. With the advent of such machine-learning methods, this hurdle may fall and hence allow to study the connection between anharmonic and non-adiabatic effects.

\section*{Acknowledgements}
This work was supported by the NOMAD Center of Excellence (European Union’s Horizon 2020 research and innovation program, Grant Agreement No. 951786), TEC1p (the European Research Council (ERC) Horizon 2020 research and innovation programme, grant agreement No. 740233), and BigMax (the Max Planck Society's Research Network on Big-Data-Driven Mater.-Science).

\end{document}